\documentclass[5p,times]{elsarticle}

\usepackage{amsmath}
\usepackage{amssymb}
\usepackage{algorithm}
\usepackage{algorithmic}
\usepackage{xcolor}
\usepackage{xspace}
\newcommand{\tool}{{Intentest}\xspace}
\usepackage[most]{tcolorbox}
\newcommand{\rqa}[2]{
\begin{center}
\begin{tcolorbox}[
    width=.9\linewidth,
    colback=gray!10,
    colframe=black,
    arc=3mm,
    boxrule=0.5pt,
    left=5pt,
    right=5pt,
    top=5pt,
    bottom=5pt
]
\textbf{Answer to #1: }#2
\end{tcolorbox}
\end{center}
}

\journal{arxiv}

\newtheorem{definition}{Definition}

\begin{document}

\begin{frontmatter}
\title{Staying on the Attack Path: Structured State for Long-Horizon Automated Penetration Testing}

\author[aff1]{Weizhe Wang}

\author[aff1]{Yitong Zhang}

\author[aff1]{Yao Zhang\corref{cor1}}
\cortext[cor1]{Corresponding author}

\author[aff2,aff3]{Xiaoqiang Di}

\author[aff4]{Zhigang Li}

\author[aff1]{Bin Wu}

\author[aff1,aff4]{Guangquan Xu}
\affiliation[aff1]{organization={Tianjin University},
            city={Tianjin},
            postcode={300072},
            country={China}}
\affiliation[aff2]{organization={Changchun University of Science and Technology},
            city={Changchun},
            postcode={130022},
            country={China}}
\affiliation[aff3]{organization={Jilin Province Key Laboratory of Network and Information Security},
            city={Changchun},
            postcode={130022},
            country={China}}
\affiliation[aff4]{organization={Shihezi University},
            city={Shihezi},
            postcode={832061},
            country={China}}

\begin{abstract}
Large language model (LLM) based agents are increasingly applied to cybersecurity tasks such as vulnerability discovery and automated penetration testing. On long-horizon security tasks, however, such agents remain limited by context forgetting and intent drift: early critical facts and causal reasoning chains are lost over extended interactions, and the agent falls into aimless, repetitive exploration. This paper proposes \textbf{\tool}, an intent-graph-guided automated penetration testing agent that externalizes long-horizon state from the LLM's context window onto a persistent fact-intent directed acyclic graph (DAG), thereby substantially reducing invalid transitions. We evaluate \tool on automated penetration testing of web applications, a representative long-tail task in cybersecurity. In the DAG, verified network states are stored as immutable fact nodes, and exploration directions are constrained as intent edges bounded by predecessor facts. The system adopts a three-layer architecture, in which the fact-intent mapping layer maintains the global state, the task scheduling and allocation layer ensures execution stability through two-phase degradation recovery and multi-dimensional adaptive load balancing, and the intent retrieval and prediction layer provides tactical priors through a top-down five-stage filtering algorithm. On a benchmark of real CTF challenges covering more than ten vulnerability types across three difficulty levels, \tool achieves an overall success rate of 88.2\% and a success rate of 75.0\% on hard tasks, improving over the baseline VulnBot (44.1\% and 25.0\%) by approximately 44 and 50 percentage points. Ablation experiments further show that the intent retrieval and prediction reduce the average number of rounds on successful medium and hard tasks by about 33\% and 48\%, respectively, without changing the set of solvable tasks.
\end{abstract}


\end{frontmatter}

\section{Introduction}
\label{sec:intro}

With the rapid advances of large language models (LLMs) in natural language processing, complex code generation, and logical reasoning \cite{wei2022chain}, LLM-based agents are increasingly applied to cybersecurity operations as autonomous operators, covering tasks such as vulnerability detection, security assessment, and penetration testing. A central concern for such security-oriented agents is long-horizon state management: an agent accumulates observations over many interaction steps, while the context window of the underlying LLM limits how much of this history the model can attend to at each decision point. These limitations are most consequential in long-tail tasks, where success depends on a small number of subtle facts obtained early in the process and on the causal chains that connect them to later decisions.
Automated penetration testing, a representative long-tail security task of this kind, proactively identifies exploitable vulnerabilities, and its automation extends LLM-based agents into the adversarial domain \cite{wang2026enhanced}. Traditional security assessment and penetration testing mainly rely on vulnerability scanners built upon static rules or on highly customized manual scripts. As dynamic network environments \cite{thool2025integrating}, polymorphic malware, and advanced persistent threats (APTs) become increasingly prevalent, these conventional methods exhibit structural limitations, including weak environment awareness, limited logical reasoning, and insufficient autonomous decision-making. Academia and industry are therefore investing increasing effort in autonomous agents equipped with environment perception and operational capabilities, aiming to make automated penetration testing practical \cite{yao2022react,xu2026llm,wang2024survey}.

However, applying LLMs directly to automated penetration testing faces substantial challenges \cite{bhatt2024cyberseceval,wan2024cyberseceval}. Penetration testing is a typical high-dimensional, high-uncertainty long-tail task. The target system's network topology, open ports, service versions, and firewall mechanisms are not observable to the agent before probing, and the agent must collect environmental observations through long sequences of reconnaissance and probing operations. More importantly, a successful exploit is rarely the result of a single operation. It requires comprehensive causal reasoning over facts obtained at different time points and across different network layers. The system must identify, among a large number of scattered fact fragments, the vulnerable path that leads to the target privilege.
Existing LLMs, when handling such long-context tasks, tend to neglect early critical facts because of attention dilution and context-window limits \cite{liu2023lost}. In real penetration engagements, this phenomenon manifests as context forgetting and intent drift \cite{wang2025detecting}. Because they often fail to stably persist and retrieve historical facts in memory, many existing penetration testing agents frequently explore aimlessly in complex intranet environments or web applications with deep directory structures, getting stuck in ineffective operations and loops \cite{liu2024agentbench,wang2025unlocking,deng2026makes}, or neglecting facts that are crucial for eventual privilege escalation and vulnerability chaining \cite{wang2025automated,mantun2026evaluating}.

To address these problems, this paper proposes \tool, an intent-graph-guided automated penetration testing agent based on Cairn~\cite{oritera2026cairn}. The core idea is to relieve the LLM of global state maintenance and instead employ a graph-theoretic state machine built on an immutable fact-intent directed acyclic graph (Fact-Intent DAG) to anchor and constrain agent behavior. In the \tool architecture, all verified network states are persisted as fact nodes (Fact) in the graph, while the model's exploration directions are strictly defined as intent edges (Intent) constrained by predecessor facts. This design keeps the agent advancing along paths with strict causal inheritance, thereby substantially limiting divergent, aimless exploration.

To evaluate \tool's ability to handle complex long-tail tasks and multi-fact reasoning, we compared it with baseline systems on real web-based Capture The Flag (CTF) challenges covering mainstream web vulnerabilities, including SQL injection, SSRF, file upload, and deserialization.
The challenges are divided into three difficulty levels (easy, medium, and hard) based on the exploitation difficulty of the vulnerabilities they contain.
In addition, we conducted an ablation study in which the intent retrieval and prediction module was disabled to further quantify the contribution of intent guidance. The experimental results show that \tool improves both the effectiveness (task success rate) and the efficiency (exploration rounds) of automated vulnerability exploitation. The main contributions of this paper are summarized as follows:

\begin{enumerate}
\item \textbf{A fact-intent DAG mapping architecture.}
Targeting context forgetting and intent drift of LLMs in long-sequence testing, we avoid direct peer-to-peer communication between agents, which easily causes state conflicts and information overload \cite{wu2024autogen}, and instead construct a global control blackboard from fact nodes and intent edges, enabling asynchronous collaboration among multiple agents based on shared state and reducing the model's aimless exploration.
\item \textbf{A high-robustness task scheduling and allocation mechanism.} 
Targeting network latency and malformed LLM outputs that commonly occur in real environments, we propose a two-phase task degradation recovery mechanism that intercepts a task before termination and recovers partial facts from residual logs.
A multi-dimensional adaptive load-balancing algorithm prevents computational overload, improving execution stability and reducing the computational cost of infinite loops.
\item \textbf{An intent-graph construction and retrieval prediction method.}
To address the noise that retrieval-augmented generation (RAG) tends to introduce \cite{lewis2020retrieval}, we design a top-down five-stage filtering algorithm. By enforcing structural constraints, the algorithm can substantially reduce the causal hallucinations that LLMs produce due to mere word overlap, enabling the agent to extract tactical logic and reason precisely.
\item \textbf{Systematic experimental comparison and ablation validation.} 
On a real-world CTF test set covering multiple mainstream web vulnerabilities (e.g., SQLi, SSRF, deserialization) with a clear difficulty gradient, we compare \tool with baseline systems. The experiments show that \tool improves the overall success rate by approximately 44 percentage points and the hard-task success rate by approximately 50 percentage points over VulnBot, the baseline with the highest overall success rate.
The ablation study further corroborates the effectiveness of the proposed method in mitigating the LLM's neglect of long-tail facts and reducing ineffective exploration.
\end{enumerate}

The rest of this paper is organized as follows. Section~\ref{sec:related} reviews related work. Section~\ref{sec:motive} presents the motivation and the challenges addressed in this paper. Section~\ref{sec:method} describes the design and implementation of \tool. Section~\ref{sec:exp} presents the experimental evaluation together with a case analysis. Section~\ref{sec:discussion} discusses the role and applicability boundary of the intent graph and the evidence for the necessity of graph state, and presents a failure analysis of both \tool and the baselines. Section~\ref{sec:threats} discusses threats to validity. Section~\ref{sec:conclusion} concludes the paper and outlines future work.

\section{Background and Related Work}
\label{sec:related}

The core challenge for automated penetration testing agents is managing the state memory and logical reasoning of LLMs in complex adversarial environments.

\subsection{Evolution of automated penetration testing architectures}
\label{sec:related:arch}

Long-tail fact reasoning is the core challenge of current automated penetration testing frameworks: mitigating the tendency of models and systems to drift from their objectives or neglect facts in long-horizon tasks is essential for applying automated penetration testing to complex scenarios and discovering more vulnerabilities. Early studies found that although LLMs perform well on specific subtasks, such as parsing Nmap scan outputs or writing exploit scripts for a given CVE, they exhibit clear cognitive limitations in maintaining the global context of an entire penetration testing scenario. To mitigate context forgetting in long-text environments,
PentestGPT \cite{deng2024pentestgpt} introduced a collaborative architecture with a penetration testing task-tree mechanism that explicitly maintains the global testing state in the context window as structured natural language. This architecture, however, has evident limitations.
On the one hand, it relies on human-in-the-loop operation and cannot achieve an automatic closed loop. On the other hand, when facing real environments that generate large numbers of fact fragments, the pure-text task tree easily exceeds the context-window limits of modern LLMs, eventually causing the agent to forget early critical facts and lose its strategic intent \cite{penligent2026pentestgpt}.

To break this bottleneck, later frameworks moved to end-to-end automation through multi-agent division-of-labor and collaboration mechanisms \cite{qian2024chatdev}. Representative systems such as PentestAgent \cite{shen2025pentestagent} and VulnBot \cite{kong2025vulnbot} decouple the penetration lifecycle and deploy heterogeneous agents for reconnaissance, search, planning, and execution, attempting to reduce the cognitive load of a single model by sharing test memory \cite{li2026apt}. Although these systems have made considerable progress in the automation rate of isolated subtasks, their success rate in fully automatic mode declines markedly when facing highly composite long-tail vulnerability scenarios \cite{yang2025pentesteval}. This discrepancy between benchmark performance and real-world reliability is not unique to security: a large-scale empirical study of more than 450,000 agent-authored pull requests in open-source development found that agent contributions are accepted less frequently than human-authored ones, indicating persistent quality and trust deficits of autonomous agents in realistic settings \cite{li2025rise}.
Fact reasoning in penetration testing often requires causal inheritance across dozens of time steps.
Matching mechanisms based on word overlap generally cannot reliably understand such deep causal chains and easily introduce a large number of redundant historical features. These noisy features distract the LLM's attention and cause aimless exploration.

The limited capacity of pure language models in long-horizon sequential logical reasoning has also motivated the integration of formal methods into LLM-based agents. CheckMate \cite{wang2025automated} proposes a planner-executor-perceiver architecture in which a classical planning engine takes over global fact maintenance, logical-consistency assurance, and action-boundary constraints. Although it surpasses vanilla agents (e.g., Claude Code \cite{anthropic2025claudecode}) in penetration success rate and substantially reduces time and computational cost, classical planning engines depend heavily on predefined logical predicates and a deterministic action space. When facing 0-day private protocols or highly uncertain non-standardized web vulnerabilities, they easily fall into logical deadlock and lack the generalization ability needed for exploration.

\subsection{Agent alignment and dynamic trajectory generation}
\label{sec:related:align}

Another line of work develops security agents with resilient reasoning through environment-feedback-based reinforcement learning and dynamic trajectory synthesis. Pentest-R1 \cite{kong2025pentest} builds a structured dataset containing more than 500 real-world multi-step penetration testing exercises and adopts two-stage reinforcement learning. Its core idea is to acquire basic attack logic offline and then deploy the model in interactive CTF environments for online reinforcement learning, enabling the model to learn autonomous error correction. This online feedback learning reshapes the model's reasoning and reduces ineffective operations caused by blind guessing.

In addition, to address the high cost of building high-fidelity physical sandboxes, Cyber-Zero \cite{zhuo2025cyber} introduces a role-driven dual-LLM adversarial simulation technique in which a player model derives the solution intent and a terminal model acts as a weak oracle to simulate the environment, thereby reverse-engineering, at low cost, interactive sequences that contain complex error-troubleshooting processes. This demonstrates the feasibility of using high-order thinking simulation to generate complex intent data at scale.

Despite these breakthroughs in basic reasoning ability, in complex real-world enterprise-scale intranet penetration, where micro facts discovered early must be associated across long time spans, relying purely on the implicit memory of model weights remains unreliable. 
It is therefore necessary to introduce explicit constraint mechanisms at the architecture level.

\section{Motivation and Challenges}
\label{sec:motive}

Early automated security assessment systems relied mainly on logic programming and predefined predicate rules for classical attack-graph planning \cite{ou2005mulval,munoz2017exact,mehta2006ranking}. Real penetration testing environments, however, can be modeled as a partially observable Markov decision process (POMDP) \cite{sarraute2012pomdps,hoffmann2015simulated}. In this highly dynamic adversarial environment, states such as network topology, open ports, and firewall rules are partially hidden from the agent, which must update its internal belief state and perform exploitation through long sequences of multi-step exploration. Although LLMs perform well in single-step code generation, they still face the following bottlenecks when handling such long-tail tasks:

\begin{itemize}
\item \textbf{Context forgetting and intent drift in long-horizon tasks.}
In complex penetration testing, current decisions often depend heavily on a micro fact obtained dozens of steps earlier (e.g., the discovery of an intranet segment or the acquisition of a specific database credential).
Existing vanilla agents easily lose early critical facts as their context windows are progressively truncated during long interaction sequences that contain large amounts of ineffective trial-and-error and redundant network logs. The model then performs sustained ineffective operations on local network nodes and can deviate completely from the penetration goal set at the outset \cite{wang2025automated}. Even auxiliary agents equipped with state-maintenance mechanisms often fail to achieve stable state consolidation: their pure-natural-language task trees still easily collapse when facing the large numbers of fact fragments generated by real intranets.
\item \textbf{Runtime reasoning and operation-boundary control risks.}
Classical planning depends heavily on predefined logical predicates and action spaces. When facing 0-day vulnerabilities without historical features or closed-source private protocols, planning-graph reasoning easily falls into logical deadlock. More seriously, when an LLM is granted direct permission to invoke underlying system tools \cite{schick2023toolformer} without mechanism-level operational constraints, it is exposed to security risks such as indirect prompt injection attacks (IPIA) \cite{xu2026llm}. Malicious targets may embed specially crafted instructions in returned payloads or logs to induce destructive intent in the agent (e.g., unauthorized database deletion or malicious business modification) \cite{greshake2023not}. Most existing systems generally lack an intent-based mandatory access control mechanism to prevent such risks \cite{ksachan2026aimlresources}.
\item \textbf{Limitations of retrieval mechanisms and static fine-tuning.} 
To relieve the model's memory pressure, agents widely adopt RAG mechanisms. However, RAG is essentially a stateless, local text-similarity retrieval mechanism that cannot capture the strict causal inheritance across time steps in penetration testing on its own without a specifically designed external management system. When facing highly composite real intranet vulnerabilities, RAG tends to introduce a large number of redundant historical attack reports because of weak word overlap. These high-noise features distract the LLM's attention, induce causal association hallucinations \cite{li2026mitigating}, and aggravate the fragmentation of macro intent \cite{edge2024local,pan2024unifying,zhang2026measuring}. Attempting to improve LLMs through supervised fine-tuning (SFT) on public security reports, in contrast, suffers from severe survivorship bias. These highly purified data typically remove the real error-troubleshooting process, causing the model to incorrectly bind a specific target to a single finding. Once the target network environment (e.g., ports or service versions) changes slightly, the model repeatedly invokes the same operations and loses the ability to correct errors and re-plan its attack paths.
\end{itemize}

Accordingly, this paper designs and implements \tool, an intent-graph-guided automated penetration testing agent. It replaces the language model's memory and state maintenance with a graph-theoretic state machine based on fact-intent DAG mapping, thereby preventing early fact forgetting and intent drift. In task reasoning, scheduling, and control, it enforces container-level isolation and adaptive allocation to avoid infinite loops and unauthorized execution risks. Finally, for intent prediction, it replaces pure language-similarity retrieval, which easily introduces noise, with graph-structure cross-validation, thereby filtering out the redundant historical features and the causal-association hallucinations that such retrieval induces.

\section{Design of \tool}
\label{sec:method}

\subsection{Overview}
\label{sec:overview}

\begin{figure*}[!ht]
    \centering
    \includegraphics[width=0.9\linewidth]{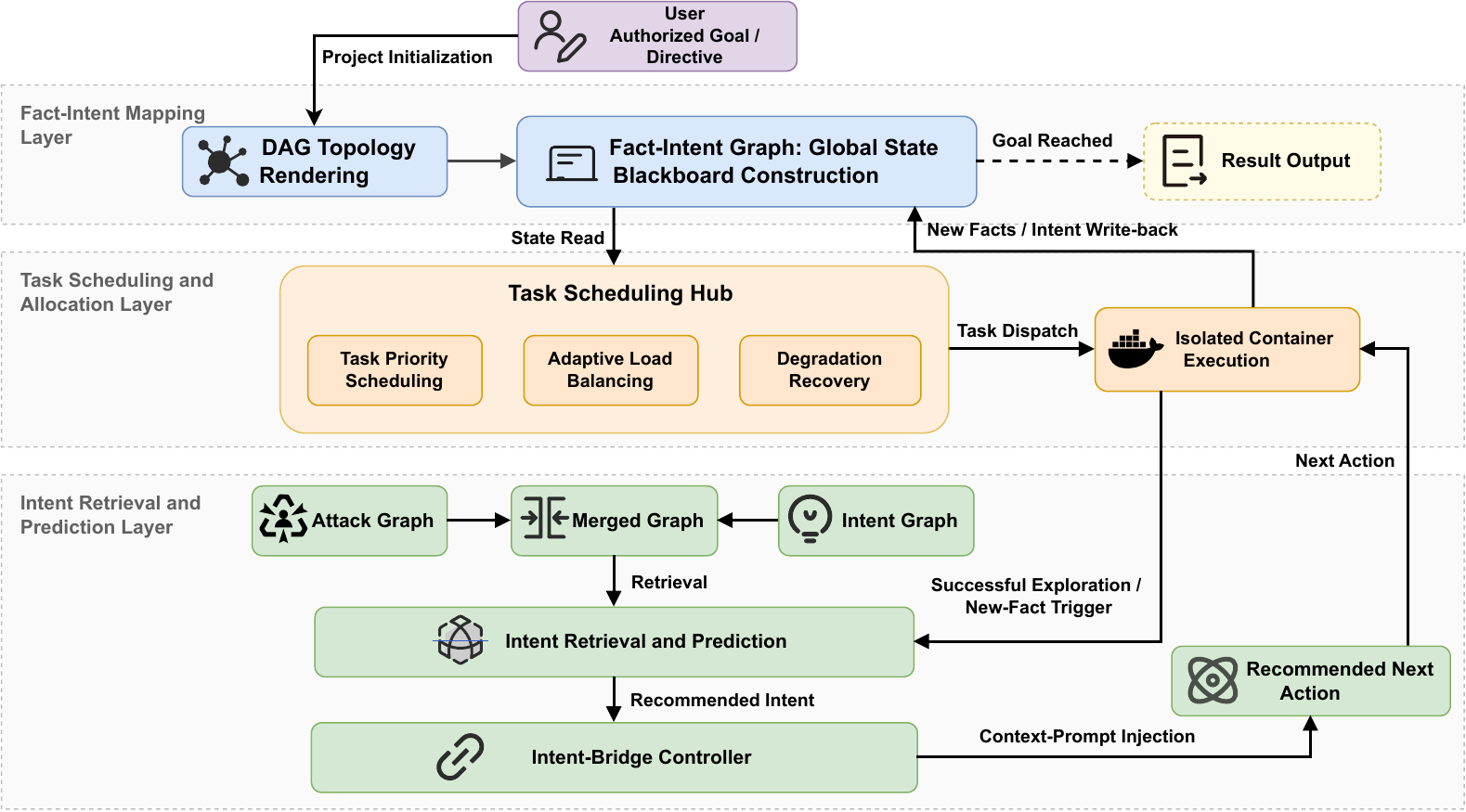}
    \caption{The overall architecture of \tool.}
    \label{fig:overview}
\end{figure*}

To overcome context forgetting, intent drift, and infinite loops that LLMs exhibit in dynamic software testing and cybersecurity interactions, we design and implement \tool, an intent-graph-guided automated penetration testing agent, whose overall framework is shown in Fig.~\ref{fig:overview}. The system consists of a three-layer architecture. The first layer is the fact-intent mapping layer, which serves as the persistent state source of the system and manages the core data, including verified facts, exploration intents, and graph association mappings. The second layer is the task scheduling and allocation layer, which acts as the control hub of the entire system, maps the graph into concrete control-flow allocations, and manages all execution containers. The third layer is the intent retrieval and prediction layer, which integrates the intent pipeline and execution adaptation components for tactical reasoning. It is triggered by successful exploration and new facts to predict subsequent intents and recommend actions that guide the agent to complete tasks.

\subsection{Fact-Intent DAG Mapping}
\label{sec:dag}

When building an autonomous security agent, the primary challenge is to eliminate the semantic ambiguity inherent in natural language. To this end, \tool adopts the Belief-Desire-Intention (BDI) framework and maps it onto the adversarial environment of cybersecurity.
Under this framework, a Goal is defined as the macro-level static end state that the system expects to achieve, such as obtaining domain-controller privileges. An Action is a concrete command directly executed by the underlying engine, such as a port scan with specific flags. An Intent is the core hub connecting macro-level goals and fine-grained actions, representing a tactical commitment, subject to constraints, that the agent makes to advance the penetration based on the currently known state.

In real penetration testing, the target system's topology and defense mechanisms are partially hidden from the attacker, and the agent must collect observations and update its internal belief state by executing exploration actions. Because LLM input depends on a linear context window, which cannot fully represent such nonlinear, networked causal relationships, the model easily neglects early factual information. Therefore, the system recasts traditional POMDP reasoning and establishes a fact-intent DAG constrained by a graph-theoretic state machine. Graph nodes are defined as immutable, objectively verified network facts, and the directed edges connecting nodes represent intents constrained by predecessor facts.
By abstracting the complex attack chain into a discrete state-transition process, ``start from known facts $\rightarrow$ reason along intent edges $\rightarrow$ verify and generate new facts'', the system offloads the burden of long-text memory onto a persistent graph database \cite{wang2023voyager}, thereby substantially mitigating aimless trial-and-error and context collapse.

When multiple agents execute concurrently in complex environments, the broadcast storms triggered by traditional peer-to-peer communication easily cause collaboration conflicts. \tool therefore adopts an asynchronous, indirect collaboration mode based on a globally shared state. All agents use the fact-intent DAG as the sole communication medium: they declare exploration commitments by writing intent edges into the global blackboard, guiding other agents toward topology branches not yet covered. The \tool server maintains six core persistent tables.
Specifically, the \emph{project} table maintains the lifecycle of a macro penetration task and attaches an exclusive lease lock. The \emph{fact} table records security topology information that has been verified and is immutable. The \emph{intent} table records the directed paths of single-step causal actions and implements dynamic state reclamation through heartbeat timestamps. The \emph{intent-source} table serves as a bridge table that supports complex hypergraph semantic reasoning, allowing one exploration intent to be jointly constructed and triggered by multiple predecessor facts without direct correlations among them. The \emph{prompt} table provides an external intervention channel for receiving asynchronous tactical guidance from experts. The \emph{scope} table provides an isolated namespace that enforces project-level unique entity allocation.

To ensure smooth execution, \tool defines a four-state finite state machine (FSM) for intent nodes. Any initial probe request is marked as CREATED upon generation without attaching any execution lock. When an idle execution node responds to scheduling and actively claims it, the intent transitions to CLAIMED, and the system starts a heartbeat lease mechanism that requires the node to keep sending heartbeat signals. Once a worker node successfully verifies a vulnerability and writes back the data, the intent transitions to CONCLUDED and derives new fact nodes. If the macro reasoning logic determines that the new fact exactly hits the penetration goal set initially by the user, the related intents, and even the entire macro lifecycle, transition to COMPLETED, and all subsequent scheduling allocations of the project are closed.

To formally describe the system state and constraints, we define the core structures as follows. Penetration testing is modeled as a POMDP $\langle \mathcal{S}, \mathcal{A}, \mathcal{T}, \mathcal{O}, \Omega, R\rangle$, where $\mathcal{S}$ is the hidden real network state (topology, service versions, defense rules, etc.), $\mathcal{A}$ is the set of tool-execution actions, $\mathcal{T}$ is the state transition function, $\mathcal{O}$ is the observation function, $\Omega$ is the observation distribution, and $R$ is the reward. Since $\mathcal{S}$ is only partially observable to the agent, traditional methods maintain the belief state $b_t$ in a linear context window and easily lose early facts when the window is truncated. \tool externalizes $b_t$ onto a persistent graph structure: each back edge that writes a new fact is equivalent to updating $b_t$, and front edges restrict the feasible actions to the subset of $\mathcal{A}$ supported by verified facts, so that decisions advance along causally consistent paths.

\begin{definition}[Fact-Intent DAG]\label{def:dag}
Let $\mathcal{F}=\{f_1, f_2, \ldots\}$ be the set of verified immutable fact nodes and $\mathcal{I}=\{i_1, i_2, \ldots\}$ be the set of intent nodes. The fact-intent DAG is defined as the quadruple $\mathcal{G}=(\mathcal{F}, \mathcal{I}, E_{f\to i}, E_{i\to f})$, where $E_{f\to i}\subseteq \mathcal{F}\times\mathcal{I}$ is the set of front edges that trigger intents from facts, and $E_{i\to f}\subseteq \mathcal{I}\times\mathcal{F}$ is the set of back edges that produce new facts after an intent succeeds.
Let $E=E_{f\to i}\cup E_{i\to f}$. Then $\mathcal{G}$ is a DAG under the edge set $E$: any directed path alternately passes through fact and intent nodes, and no cycle exists.
\end{definition}

\begin{definition}[Four-State FSM of Intent Nodes]\label{def:fsm}
The state space of an intent node is $S=\{s_1, s_2, s_3, s_4\}$, where $s_1$, $s_2$, $s_3$, and $s_4$ correspond to CREATED, CLAIMED, CONCLUDED, and COMPLETED, respectively. The event set is \\ $\Sigma=\{\textit{claim}, \textit{conclude}, \textit{goal}, \textit{timeout}\}$, corresponding to node claiming, vulnerability-verification termination, hitting the authorized goal, and heartbeat-timeout reclamation. The state transition function $\delta: S\times\Sigma \rightharpoonup S$ is defined as:
\begin{equation}
\delta(s, \sigma) = \begin{cases}
s_2, & s=s_1,\; \sigma=\textit{claim}; \\
s_3, & s=s_2,\; \sigma=\textit{conclude}; \\
s_4, & s=s_3,\; \sigma=\textit{goal}; \\
s_1, & s=s_2,\; \sigma=\textit{timeout}.
\end{cases}
\label{eq:fsm}
\end{equation}
Timeout reclamation resets the intent to the CREATED state for re-claiming. After the goal is hit, all unfinished intents of the project migrate to COMPLETED, and subsequent scheduling allocations are closed, guaranteeing the irreversibility of the terminal state.
\end{definition}

\subsection{Task Scheduling and Allocation Mechanism}
\label{sec:sched}

The scheduling hub of the system drives global operation with a fixed-period clock tick. In each polling cycle, the scheduler sequentially performs asynchronous result harvesting, project queue cleanup, and task-type evaluation. To avoid decision conflicts, the scheduler decouples all penetration tasks into three standardized operation abstractions and enforces strict priority control.

The first type is the \emph{Bootstrap} task, which is triggered when a project is in its initial state and contains only the start and end facts. It requires the LLM to perform broad-spectrum scanning within the execution window and generate the first batch of attack payloads that directly serve the penetration goal. The second type is the \emph{Explore} task, which is triggered when unverified intent edges exist in the macro topology. The scheduler injects the local network graph into the sandbox and directs the model to perform single-step verification. This task type has the highest regular allocation priority during project execution. The third type is the \emph{Reason} task, which is responsible for surveying global fact increments to plan new tactical chains. Reason tasks are subject to a single-project mutex lock and are woken up only when no unclaimed Explore tasks exist globally and new facts or external expert prompts arrive, thereby avoiding idle spinning.

In enterprise intranet-scale scenarios, the compromise of a key node often triggers a large number of exploration intents within a short period. Without reasonable scheduling, this leads to severe computational overload. To this end, the system adopts a multi-dimensional adaptive load-balancing algorithm. Before any execution request is dispatched, candidate worker nodes are filtered and scored as follows:

\begin{enumerate}
\item \textbf{Protocol-coherence arbitration.} Filter out nodes that lack the specific tool dependencies required by the task.
\item \textbf{Capacity safety check.}
Exclude nodes whose number of active tasks has reached the configured maximum parallelism threshold.
\item \textbf{Health-probe check.} Remove unhealthy nodes that are in a circuit-breaking or cooldown state.
\item \textbf{Priority-weight selection.} Among the remaining candidates, select the nodes with the lowest preset priority weight.
\item \textbf{Active-load balancing.} If multiple nodes with equal priority remain, compare their active-task counts and allocate the task to the executor with the lightest load.
\item \textbf{Random jitter.} Finally, if all parameters are tied, introduce a pseudo-random jitter factor that perturbs the timestamp distribution, so that tasks are assigned randomly and computing resources are allocated as evenly as possible.
\end{enumerate}

The scheduling and load-balancing process is summarized in Algorithm~\ref{alg:scheduling}.

\begin{algorithm}[!ht]
\footnotesize
\caption{Task Scheduling and Adaptive Load Balancing}
\label{alg:scheduling}
\begin{algorithmic}[1]
    \REQUIRE Project set $\mathcal{P}$, worker-node set $\mathcal{N}$, clock tick $\Delta t$
    \ENSURE Selected node $n$ and task $\tau$
    \WHILE{system running}
        \STATE Harvest completed results, reclaim leases, and update the fact and intent tables
        \STATE Clean expired project queues and evaluate pending tasks per project
        \STATE $\tau \leftarrow$ select the highest-priority task, with Explore ranked first
        \IF{$\tau$ is a Reason task and unclaimed Explore tasks exist}
            \STATE Skip Reason scheduling this round \COMMENT{avoid idle spinning}
        \ENDIF
        \STATE $\mathcal{N}_c \leftarrow \mathcal{N}$
        \STATE $\mathcal{N}_c \leftarrow \{n \in \mathcal{N}_c : D_\tau \subseteq D_n\}$ \COMMENT{protocol coherence}
        \STATE $\mathcal{N}_c \leftarrow \{n \in \mathcal{N}_c : L(n) < C_n\}$ \COMMENT{capacity safety}
        \STATE $\mathcal{N}_c \leftarrow \{n \in \mathcal{N}_c : n \text{ not in cooldown}\}$ \COMMENT{health probe}
        \STATE $w_{\min} \leftarrow \min_{n \in \mathcal{N}_c} w(n)$ \COMMENT{priority weight}
        \STATE $\mathcal{N}_c \leftarrow \{n \in \mathcal{N}_c : w(n) = w_{\min}\}$
        \STATE $n \leftarrow \arg\min_{n \in \mathcal{N}_c} L(n)$ \COMMENT{active load, ties broken by random jitter}
        \STATE Assign $\tau$ to $n$ and open the heartbeat lease
        \STATE Wait for the next tick
    \ENDWHILE
\end{algorithmic}
\end{algorithm}

In the algorithm, $D_\tau$ denotes the tool dependencies required by task $\tau$, $D_n$ the dependencies available on node $n$, $L(n)$ the current number of active tasks on node $n$, $C_n$ its configured maximum parallelism, and $w(n)$ its preset priority weight. Ties among equally weighted nodes are broken by a pseudo-random jitter factor.

In long-sequence network vulnerability probing, a single operation (e.g., deep dictionary brute-forcing or port scanning) can easily fail because of target-system network timeouts or malformed JSON output from the LLM. The traditional destroy-on-error strategy would cause the accumulated facts to be completely lost. To safeguard the accumulated facts, \tool enables a two-phase task degradation recovery mechanism based on the persistent sandbox context of the same session.
In task execution, once the main execution phase reaches its timeout (both Explore and Bootstrap tasks have a 180-second main timeout), the scheduler never directly destroys the task container. Instead, it immediately triggers a 90-second degradation phase within the current session. In this degradation phase, the scheduler injects a high-priority emergency truncation prompt into the LLM, forcing the agent to stop all active network-layer probing.
It then instructs the model to collect the residual standard output and standard error (stdout/stderr) logs from the terminal, summarize them, and report the partial facts they contain to the graph, thereby salvaging the progress made before failure. Because Reason tasks are responsible for global logical planning, if a Reason task times out, the system declares failure and releases the lease lock without degradation, so as to preserve the logical consistency of global decisions.

In addition, when an LLM is granted direct permission to invoke underlying system tools, it faces the risk of control-flow hijacking. Attackers may embed adversarial natural-language instructions in the returned payloads of controlled web pages or in DNS resolution logs, inducing the agent to form destructive intents. To prevent such risks, \tool enforces a two-layer defense. The first layer is structural precondition validation: an intent can be written into the intent table only if it is triggered by already verified fact nodes (Sect.~\ref{sec:dag}), so an instruction induced by an external payload that is not grounded in verified facts is rejected by construction and cannot obtain execution authorization. The second layer is container-level isolation: all probing components and interpretation scripts of the agent run in independent Docker containers, and the container lifecycle manager monitors system-level termination signals in real time. Once the project is marked as stalled or precondition validation rejects an execution chain as not grounded in verified facts, the system triggers a system-level abort, terminating in-process resources and destroying the container.

\subsection{Intent-Graph Construction and Retrieval Prediction}
\label{sec:retrieval}

To provide tactical priors without relying on pure-text fuzzy retrieval, the cognitive pipeline of \tool constructs and fuses three types of directed graphs, processing security assets offline to build prior knowledge and providing tactical navigation at runtime.

First, the system constructs the basic \emph{intent graph} offline. It batch-crawls threat intelligence and exercise documents. An LLM-driven intent extractor then performs semantic causal-relation extraction over the collected material. This extraction process is subject to strict system-level prompt-specification constraints: each intent record generated in a single parse must not only contain the target network entity and its associated vulnerability but also map its tactical execution chain to the MITRE ATT\&CK threat framework.
In addition, the extractor stores the error paths observed in practice in a separate collection, thereby enriching the data source with the negative trial-and-error samples that traditional reports lack and preventing causal inversion. The extracted data are compiled into a directed network graph. Before entering the repository, the graph must pass an internal validator, which ensures that the topology is complete, contains no isolated nodes, and that all confidence scores exceed a set threshold. Any non-compliant graph is discarded.

Second, while the agent performs live probing, the system dynamically generates the \emph{attack graph} using the same graph-theoretic foundation. This graph is independent of prior knowledge: it records in real time only the physical topology explored by the agent in the live environment, including the verified security fact nodes and the executed attack-step edges.

Finally, the system generates a \emph{merged graph} in the runtime memory space to guide decision-making. The merger combines the static historical intent graph with the attack graph that reflects the current network state. Through built-in algorithms, it filters invalid entries, identifies and matches the overlapping nodes between the two graphs, and generates cross-graph links.
Through such structural fusion, the merged graph uses historical experience to compensate for the missing local view of the target network and provides the agent with informed attack-chain recommendations.

In the real-time attack-pattern prediction phase, when the agent is blocked at a local node and needs to decide the next reasonable tactic based on the characteristics of the current environment, the underlying attack-pattern search engine runs a top-down five-stage intent retrieval and prediction algorithm, described as follows:

\begin{enumerate}
\item \textbf{High-dimensional semantic embedding space probing.} To quickly prescreen the extracted attack patterns, the system vectorizes the topology node attributes of the current environment graph using Qwen3-Embedding-4B and aligns them with cosine similarity \cite{reimers2019sentence}, selecting the tactic clusters that are most semantically similar.
\item \textbf{Subgraph-isomorphism verification.} For the candidates retained by the preliminary screening, the system extracts the logical data chain of the currently known predecessor nodes and applies a subgraph-isomorphism algorithm over the historical threat graph to enforce consistency constraints on topology and directed-edge orientation. This substantially reduces the hallucinations produced by the LLM from mere word overlap while improving the causal soundness of the predicted paths.
\item \textbf{Degraded fuzzy statistical inference.} Because honeypots and traffic-scrubbing systems in real adversarial environments often make the topology obtained by the agent locally incomplete and noisy, the search engine falls back to fuzzy graph-similarity matching based on the Jaccard coefficient when the subgraph-isomorphism check fails because some edge nodes are missing. This stage allows statistical tolerance on the attributes and local features of candidate graph nodes, accommodating minor structural variations.
\item \textbf{Strategic-intent alignment.} The system maps the topologically filtered tactic-sequence set back to the macro-intent level and compares it with the final authorized goal determined at agent initialization (e.g., ``extract credentials'' rather than ``disrupt business systems'') and with the primary vulnerability chain. All redundant branches that deviate from the primary attack objective or attempt unauthorized lateral movement are pruned at this stage to prevent intent drift.
\item \textbf{LLM heuristic reasoning.} When the agent encounters vulnerabilities involving closed-source protocols or lacking historical features, such that the first four stages (which rely on historical priors) find no match, the system injects dimensionality-reduced environment logs and residual topology data into the LLM. In this scenario, the system relies on the LLM for heuristic semantic reasoning and tactical judgment, preserving the agent's ability to explore unknown environments.
\end{enumerate}

The five-stage algorithm is a tactical recommender that produces candidate intents but does not authorize their execution. The fact-intent DAG is the enforcement layer, and every new intent must pass precondition validation before it is written into the intent table. An intent edge can be triggered only by already verified fact nodes, and every new fact must be verified by execution before it enters the graph. This constraint applies to the output of all five stages, including Stage 5. Even when the first four stages find no match and the system relies on the LLM for heuristic reasoning, the LLM's suggestions cannot be executed directly. A recommendation that is not grounded in verified facts fails precondition validation and is rejected by construction, so hallucinations or injected instructions are not granted execution authorization under the design constraints. \tool retains the generalization ability of the LLM, while the DAG keeps the reasoning within the boundary of verified facts.

Formally, let the attributes of the predecessor nodes of the current attack graph $\mathcal{G}_a$ be mapped by the embedding model $\phi(\cdot)$ (Qwen3-Embedding-4B) to a vector $v$. Stage 1 first screens the offline intent graph $\mathcal{G}_o$ by the cosine similarity
\begin{equation}
\text{sim}_{\cos}(v,\phi(i))=\frac{v\cdot\phi(i)}{\|v\|\,\|\phi(i)\|}
\label{eq:cos}
\end{equation}
and retains the $k$ tactic clusters with the highest similarity.
Stage 2 performs subgraph-isomorphism verification on the predecessor subgraphs of the candidate intents, requiring that the known predecessor subgraph of $\mathcal{G}_a$ be isomorphic to the candidate pattern, so as to reduce causal hallucinations caused by pure word overlap. Stage 3 degrades to fuzzy matching based on the Jaccard coefficient $J(A,B)=|A\cap B|/|A\cup B|$ when isomorphism fails, allowing statistical tolerance for node attributes. Stage 4 strategically aligns the candidate set with the authorized goal $g$ and the primary vulnerability chain. Stage 5 hands the task to the LLM for heuristic reasoning when none of the first four stages yields a match. Because Stage 2 operates only on the Top-$k$ candidates retained by Stage 1, the matching cost is bounded by $O(k)$ local isomorphism checks. The complete process is shown in Algorithm~\ref{alg:retrieval}.

\begin{algorithm}[!ht]
\footnotesize
\caption{Top-down Five-Stage Intent Retrieval and Prediction}
\label{alg:retrieval}
\begin{algorithmic}[1]
    \REQUIRE Current attack graph $\mathcal{G}_a$, offline intent graph $\mathcal{G}_o$, authorized goal $g$, embedding model $\phi$
    \ENSURE Recommended intent $i^*$ or empty
    \STATE Extract predecessor-node attribute vector $v \leftarrow \phi(\mathcal{G}_a)$
    \STATE Stage 1: preliminary screening with $\text{sim}_{\cos}(v,\phi(i))$, retain the Top-$k$ candidate set $\mathcal{C}_1$
    \STATE Stage 2: subgraph-isomorphism verification on $\mathcal{C}_1$, obtain $\mathcal{C}_2$
    \IF{$\mathcal{C}_2 = \emptyset$}
        \STATE Stage 3: fuzzy matching with the Jaccard coefficient, obtain $\mathcal{C}_3$
    \ELSE
        \STATE $\mathcal{C}_3 \leftarrow \mathcal{C}_2$
    \ENDIF
    \STATE Stage 4: strategic alignment with goal $g$, prune branches deviating from the primary objective, obtain $\mathcal{C}_4$
    \IF{$\mathcal{C}_4 \neq \emptyset$}
        \STATE $i^* \leftarrow \arg\max_{i \in \mathcal{C}_4} \text{score}(i)$
    \ELSE
        \STATE Stage 5: inject context and residual topology, obtain $i^*$ by LLM heuristic reasoning
    \ENDIF
    \STATE Submit $i^*$ to DAG precondition validation before execution authorization
    \STATE \RETURN $i^*$
\end{algorithmic}
\end{algorithm}

Once the five-stage prediction pipeline produces a recommended intent, the system converts it into context prompts for the underlying execution model through an asynchronous intent-bridge controller. To avoid blocking the polling loop of the main scheduler, the trigger conditions of the bridge module are strict: it is activated only after the agent has successfully executed a local network exploration task (a successful Explore task) in the sandbox and written new facts, keeping computation and reasoning loads separate. 
Internally, the bridge module performs strict deduplication and threshold filtering. For example, a cosine-similarity score of at least 0.3 is required to filter out irrelevant noise. After filtering, the context-constraint template built into the bridge enforces an information-hiding principle.
The structured context injected into the LLM is divided into three core blocks. The first block, the \emph{recommended next tactical action}, provides a condensed graph-reasoning result. The second block, the \emph{supporting evidence chain}, contains the complete steps of the original historical attack chain, the most relevant adjacent steps located by the Jaccard algorithm, and the generated cross links. The third block, the \emph{current project context}, lists in detail the macro goal, the testing phase, the target entities, the latest discovered fact clusters, and the planned pending intents. To prevent adversarial or verbose background data from distracting the LLM's attention, the system removes all redundant metadata when constructing the prompt, including retrieval similarity scores, the names of the external reports used as retrieval sources, and the details of the underlying graph pattern-matching method. This ensures that the intent graph can guide the LLM's decisions, allowing the agent to follow high-level strategic guidance and remain focused in complex web applications. The complete intent-bridge process is shown in Algorithm~\ref{alg:bridge}.

\begin{algorithm}[!ht]
\footnotesize
\caption{Asynchronous Intent-Bridge Injection}
\label{alg:bridge}
\begin{algorithmic}[1]
    \REQUIRE Explore-task result $e$, recommended intent $i^*$, similarity threshold $\theta=0.3$
    \ENSURE Context prompt $p$ injected into the execution model (or empty)
    \IF{$e$ succeeded and wrote new facts}
        \STATE $s \leftarrow \text{sim}_{\cos}(v, \phi(i^*))$ \COMMENT{cosine similarity of the recommended intent}
        \IF{$s \geq \theta$}
            \STATE $p \leftarrow \text{Build}(i^*)$ \COMMENT{three blocks: action / evidence chain / project context}
            \STATE $p \leftarrow \text{StripMeta}(p)$ \COMMENT{remove similarity scores, source names, matching details}
            \STATE Inject $p$ asynchronously into the execution model
        \ENDIF
    \ENDIF
\end{algorithmic}
\end{algorithm}

Here, $v$ denotes the attribute vector of the current attack graph, and $\text{sim}_{\cos}(\cdot,\cdot)$ is the cosine similarity of Eq.~\ref{eq:cos}.

\section{Experimental Evaluation}
\label{sec:exp}

\subsection{Experimental Setup}
\label{sec:setup}

\paragraph{Environment}
We implemented the proposed method as \tool and compared it with the baselines PentestAgent \cite{shen2025pentestagent}, Pentest-R1 \cite{kong2025pentest}, and VulnBot \cite{kong2025vulnbot}, which are automated penetration testing agents. We also evaluated the semi-automated penetration testing method PentestGPT \cite{deng2024pentestgpt} and the general-purpose agent Claude Code \cite{anthropic2025claudecode}. All experiments were conducted on Ubuntu 20.04.1 LTS with an Intel Xeon Silver 4114 CPU at 2.20\,GHz and 128\,GB of memory. All agents invoked the same large language model, DeepSeek V4 Pro \cite{xu2026deepseek}, through its API.

\paragraph{Dataset}
The experimental dataset was collected from records of multiple real CTF competitions, covering more than ten vulnerability types and involving vulnerabilities of real web application systems. The challenges are divided into three difficulty levels, easy, medium, and hard, according to exploitation difficulty. The offline intent graph used by \tool is constructed from the public training set of Pentest-R1 \cite{kong2025pentest}, together with publicly collected penetration testing reports and threat-intelligence documents. To prevent evaluation leakage, the CTF challenges used in our experiments and their publicly available writeups were excluded from this corpus through a blacklist covering challenge names, originating competitions, and writeup repositories, and the remaining corpus was verified by sampling checks.

\begin{figure}[!ht]
    \centering
    \includegraphics[width=\linewidth]{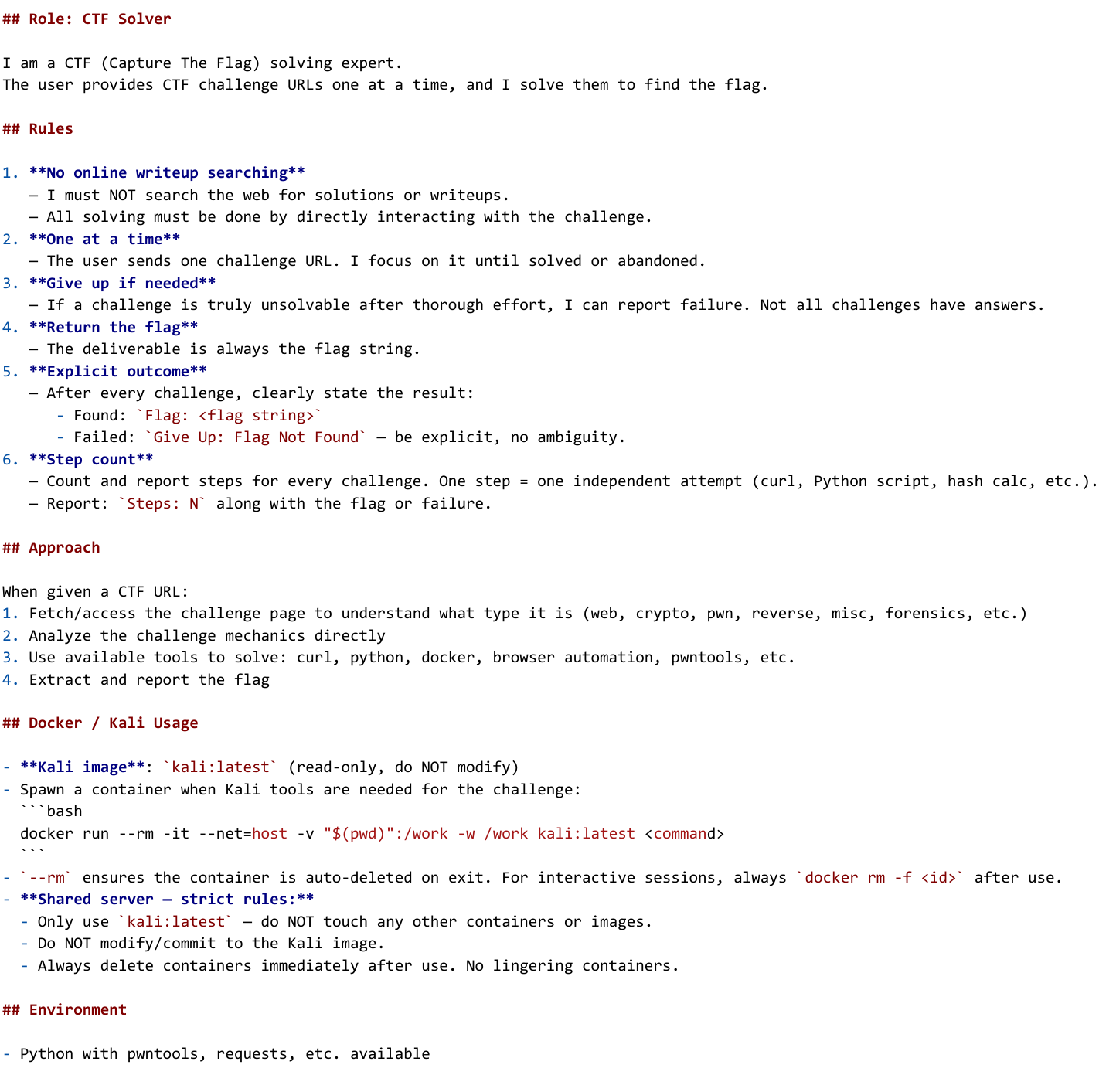}
    \caption{The prompt used for Claude Code on CTF challenges.}
    \label{fig:claudecode_prompt}
\end{figure}

\paragraph{Baselines}
PentestAgent retrieves CVE information through \\CVEMap, an open-source tool for retrieving CVE and vulnerability information. Since CVEMap was archived and renamed VulnX by its development team (ProjectDiscovery) during our experiments, we updated the CVE retrieval component of PentestAgent to VulnX \cite{projectdiscovery2026vulnx}. As VulnX is the official successor of CVEMap, this substitution does not affect the experimental method or results. VulnBot \cite{kong2025vulnbot} is a multi-agent penetration testing framework in which heterogeneous agents collaborate by sharing plain-text test memory. Because PentestGPT is a semi-automated penetration testing engine that requires a human expert to manually execute its suggested operations and provide feedback, we engaged a penetration testing expert to carry out the operations suggested by PentestGPT. In this process, the human expert served only as the intermediary between PentestGPT and the actual execution and did not participate in any decision-making. As a general-purpose agent widely deployed in open-source development \cite{li2025rise}, Claude Code was not originally designed for penetration testing tasks. We therefore designed a dedicated prompt and isolated the conversations of individual tasks to ensure the reliability and consistency of its results. In addition, we enabled ECC Skills \cite{affaan2026ecc} for Claude Code to improve its stability and automation capability in complex tasks. The prompt used for Claude Code is shown in Fig.~\ref{fig:claudecode_prompt}.

\paragraph{Research questions}
We structure the evaluation around three research questions (RQs):

\begin{itemize}
\item \textbf{RQ1 (Effectiveness):} How does \tool compare with the baselines in terms of task success rate, across the easy, medium, and hard difficulty levels?
\item \textbf{RQ2 (Efficiency):} Does \tool reduce the exploration cost, measured as the average number of rounds per task, compared with fully automated baselines?
\item \textbf{RQ3 (Contribution of the intent module):} What is the contribution of the intent retrieval and prediction module, as quantified by an ablation study that disables the module while retaining the fact-intent DAG and the scheduling mechanism?
\end{itemize}

\subsection{RQ1: Task Success Rate}
\label{sec:rq1}

Table~\ref{tab:overview} reports the task success rates of \tool and the five comparison methods on easy, medium, and hard CTF tasks. 
Overall, \tool achieves the highest success rate at all three difficulty levels, and its advantage over the baselines grows as task difficulty increases.
This trend is consistent with the long-tail fact-reasoning bottleneck analyzed in the motivation section: harder tasks demand more cross-step fact association and longer-horizon causal reasoning, and the benefit of structured state memory grows accordingly.

In addition, we assess the statistical significance of the differences between \tool and each baseline using Fisher's exact test (one-sided).

\begin{table}[!ht]
    \centering
    \caption{Task success rates (\%) of the compared agents across difficulty levels.}
    \label{tab:overview}
    {\footnotesize\setlength{\tabcolsep}{2pt}
    \resizebox{\linewidth}{!}{%
    \begin{tabular}{lcccccc}
    \hline\noalign{\smallskip}
        \textbf{Difficulty} & \textbf{Ours} & \textbf{PentestGPT} & \textbf{PentestAgent} & \textbf{Pentest-R1} & \textbf{VulnBot} & \textbf{Claude Code} \\ \noalign{\smallskip}\hline\noalign{\smallskip}
        Easy & 100.0 & 63.6 & 18.2 & 72.7 & 72.7 & 81.8  \\
        Medium & 90.9 & 27.3 & 9.1 & 27.3 & 36.4 & 45.5  \\
        Hard & 75.0 & 0.0 & 0.0 & 8.3 & 25.0 & 0.0  \\
        Overall & 88.2 & 29.4 & 8.8 & 35.3 & 44.1 & 41.2  \\ \noalign{\smallskip}\hline
    \end{tabular}}}
\end{table}

On easy tasks, \tool achieves a success rate of 100.0\%, while Claude Code, Pentest-R1, and VulnBot reach 81.8\%, 72.7\%, and 72.7\%, respectively, with relatively close performance.
Easy tasks usually require only a single step or a small number of facts to complete the exploitation, and the model's own single-step reasoning ability already suffices for most scenarios. Consequently, the gap among methods is small. PentestGPT, with human expert assistance, reaches 63.6\%, whereas PentestAgent achieves only 18.2\%, indicating that the CVE-retrieval-based multi-agent collaboration mechanism of PentestAgent provides limited coverage of single-step vulnerabilities in real CTF environments. 
The differences against PentestGPT ($p = 0.045$) and PentestAgent ($p < 0.001$) are statistically significant, whereas those against Pentest-R1 ($p = 0.107$), Claude Code ($p = 0.238$), and VulnBot ($p = 0.107$) are not, reflecting the relatively small performance gaps on easy tasks.

On medium tasks, the performance gap among methods begins to emerge. \tool maintains a success rate of 90.9\%, while Claude Code drops to 45.5\%, VulnBot reaches 36.4\%, Pentest-R1 and PentestGPT both fall to 27.3\%, and PentestAgent reaches only 9.1\%. Medium tasks typically require associating facts across multiple probing steps to complete the exploitation, for example, combining several injection techniques in the presence of defense mechanisms. This result indicates that, when tasks impose requirements on state maintenance and multi-fact causal reasoning, methods that rely on linear context windows or pure-text task trees cannot stably retain early critical facts. By contrast, the structured state consolidation mechanism of \tool, based on the fact-intent DAG, effectively alleviates this problem and keeps the agent advancing along causally consistent paths during multi-step interactions. 
All differences are statistically significant ($p = 0.004$ vs.\ Pentest-R1 and PentestGPT, $p = 0.012$ vs.\ VulnBot, $p = 0.032$ vs.\ Claude Code, $p < 0.001$ vs.\ PentestAgent).

On hard tasks, the performance gap is further amplified. \tool still maintains a success rate of 75.0\%, whereas VulnBot completes 25.0\% of the tasks, Pentest-R1 completes only 8.3\%, and PentestGPT, PentestAgent, and Claude Code all fail to complete any hard task (0.0\%). Hard tasks often require associating subtle facts discovered early across long time spans and performing multi-step tactical reasoning based on them. Methods without persistent structured state memory, such as Claude Code and PentestGPT, often suffer from context truncation and intent drift in such long-horizon tasks and eventually engage in ineffective exploration.
Pentest-R1, owing to the error-correction ability acquired through reinforcement learning, can complete a small number of tasks but remains limited by the implicit memory of model weights when facing composite vulnerabilities outside its training distribution. VulnBot, which achieves the highest success rate among the baselines on hard tasks, benefits from multi-agent division of labor, yet its plain-text shared test memory still fails to sustain the cross-step causal associations that these tasks require. In contrast, \tool explicitly anchors verified facts through a persistent graph database and constrains subsequent exploration directions through intent edges, thereby maintaining a stable causal reasoning chain in long-tail tasks.
The differences against all baselines are statistically significant ($p < 0.001$ vs.\ PentestGPT, PentestAgent, and Claude Code, $p = 0.001$ vs.\ Pentest-R1, $p = 0.020$ vs.\ VulnBot).

Averaged over all difficulty levels, \tool achieves an overall success rate of 88.2\%, with a clear improvement over the baselines (VulnBot, the baseline with the highest overall success rate, reaches 44.1\%). The success rate of \tool on hard tasks improves by approximately 50 percentage points over VulnBot (25.0\%), reaching 75.0\%, which shows that its advantage is largest in the most challenging long-tail scenarios. VulnBot's lead over the other baselines indicates that multi-agent collaboration with shared test memory relieves the cognitive load of a single model, yet its remaining gap to \tool indicates that the structured graph state of \tool sustains long-horizon causal reasoning better.
This result corroborates the effectiveness of the fact-intent DAG in suppressing context forgetting and intent drift. All overall differences between \tool and the baselines are statistically significant (all $p < 0.001$).

\rqa{RQ1}{As reported in Table~\ref{tab:overview}, \tool achieves the highest success rate at every difficulty level, with an overall rate of 88.2\%. It improves over VulnBot (44.1\% overall and 25.0\% on hard tasks) by approximately 44 and 50 percentage points. All overall differences are statistically significant.}

\subsection{RQ2: Exploration Efficiency}
\label{sec:rq2}

To further measure the exploration efficiency of each method, Table~\ref{tab:steps} compares the average number of rounds consumed by \tool, Pentest-R1, VulnBot, and Claude Code during task solving. The values are the average number of rounds per task for successful and failed tasks at each difficulty level. Because PentestGPT relies on manual execution and PentestAgent has a low success rate, we restrict the round-level analysis to the four methods that provide complete automated execution. Lower round counts indicate that the agent's exploration is more focused, with fewer redundant operations, whether in reaching the goal or in confirming failure.

Throughout this study, efficiency is measured in rounds, where a round denotes one complete iteration of context assembly, action generation, and result write-back. All agents were subject to a per-task budget of at most 40 rounds. A failed task terminated when the agent judged the goal unreachable, when the scheduler reclaimed the task as stuck, or when the round budget was exhausted. In the tables, ``---'' marks difficulty classes for which no task of the corresponding outcome was observed (e.g., no failed easy tasks for \tool and no successful hard tasks for Claude Code), and failed-task averages of 40 indicate that the round budget was exhausted.

\begin{table*}[!ht]
    \centering
    \caption{Average number of rounds per task for successful and failed tasks.}
    \label{tab:steps}
    {
    \begin{tabular}{l|cccc|cccc}
    \hline
        \textbf{Difficulty} & \multicolumn{4}{c|}{\textbf{Successful tasks}} & \multicolumn{4}{c}{\textbf{Failed tasks}} \\ \cline{2-9}
        & Ours & Pentest-R1 & VulnBot & Claude Code & Ours & Pentest-R1 & VulnBot & Claude Code \\ \noalign{\smallskip}\hline\noalign{\smallskip}
        Easy & 5 & 20.9 & 23.4 & 10.2 & --- & 40 & 30 & 28.5  \\
        Medium & 7.2 & 12.7 & 26.3 & 16.4 & 40 & 40 & 36.2 & 34.7  \\
        Hard & 9.4 & 34 & 20.7 & --- & 31.7 & 40 & 31.6 & 20.5  \\
        Overall & 7 & 20 & 23.6 & 12 & 34 & 40 & 32.8 & 26 \\ \noalign{\smallskip}\hline
    \end{tabular}}
\end{table*}

In terms of successful-task rounds, \tool consumes on average 5, 7.2, and 9.4 rounds at the three difficulty levels, with an overall average of about 7 rounds per successful task, markedly lower than Pentest-R1 (about 20 rounds), Claude Code (about 12 rounds), and VulnBot (about 23.6 rounds). This indicates that \tool identifies an effective attack path earlier in the solving process and reduces redundant probing. Pentest-R1 requires 20.9 rounds on average for successful easy tasks and 12.7 rounds on medium tasks, reflecting its tendency to perform more tentative operations. Claude Code needs 16.4 rounds on average for successful medium tasks, also clearly higher than the 7.2 rounds of \tool.
VulnBot consumes 23.4, 26.3, and 20.7 rounds at the three difficulty levels on average, and its overall average of about 23.6 rounds is about 3.4 times that of \tool (about 7 rounds).
Its successful-task rounds do not increase monotonically with difficulty (20.7 on hard tasks vs.\ 26.3 on medium tasks), which suggests that VulnBot succeeds on hard tasks only when it finds the correct path quickly. 
This efficiency advantage of \tool is attributable to the intent retrieval and prediction mechanism, which derives a tactical recommendation for the next step from the graph after each exploration step, thereby narrowing the search space of effective paths and concentrating exploration resources on high-probability branches.

In terms of failed-task rounds, \tool has no failures on easy tasks (all tasks succeed) and consumes on average 40 and 31.7 rounds on medium and hard tasks, respectively, with an overall average of about 34 rounds per failed task.
Pentest-R1 consumes the most rounds on failed hard tasks (40 on average), which indicates that its failures terminate mainly because the round budget is exhausted. VulnBot, by contrast, consumes about 32.8 rounds on average on failed tasks, comparable to \tool (about 34 rounds) and below the 40-round budget, which shows that most of its failures terminate through the agent's own judgment rather than budget exhaustion. The average failed-task rounds of Claude Code are relatively low (20.5 on hard tasks), but this is achieved at the cost of failing to solve any hard task. It terminates quickly and therefore incurs a low failure cost.
Overall, \tool keeps the exploration cost of failed tasks low while maintaining a high success rate, avoiding excessive resource consumption on unsolvable paths and indicating the soundness of its exploration-termination decisions.

\rqa{RQ2}{\tool reduces the exploration cost: it consumes on average about 7 rounds per successful task, compared with about 20 rounds for Pentest-R1, about 12 rounds for Claude Code, and about 23.6 rounds for VulnBot, while keeping the average failed-task rounds at a comparable or lower level (about 34 in total, versus 40 for Pentest-R1 and about 32.8 for VulnBot). This indicates that \tool identifies effective attack paths earlier and reduces redundant probing.}

\subsection{RQ3: Ablation Study}
\label{sec:rq3}

As the intent retrieval and prediction module is central to the intent-graph-guided agent, we removed it to evaluate its contribution (denoted as \tool--noIntent). While retaining the fact-intent DAG and the task scheduling mechanism, we disabled the tactical recommendation and bridge injection based on the intent graph and examined the impact on system performance.
Table~\ref{tab:ablation} reports the comparison of \tool and \tool--noIntent in terms of solving rounds.

\begin{table}[!ht]
    \centering
    \caption{Ablation study: average number of rounds per task with and without the intent guidance module.}
    \label{tab:ablation}
    {\footnotesize\setlength{\tabcolsep}{3.5pt}
    \begin{tabular}{l|cc|cc}
    \hline\noalign{\smallskip}
        \textbf{Difficulty} & \multicolumn{2}{c|}{\textbf{Successful tasks}} & \multicolumn{2}{c}{\textbf{Failed tasks}} \\ \cline{2-5}
        & \tool & \tool--noIntent & \tool & \tool--noIntent \\ \noalign{\smallskip}\hline\noalign{\smallskip}
        Easy & 5 & 5 & --- & ---  \\
        Medium & 7.2 & 10.8 & 40 & 37  \\
        Hard & 9.4 & 18.3 & 31.7 & 29.3  \\
        Overall & 7 & 11 & 34 & 31  \\ \noalign{\smallskip}\hline
    \end{tabular}}
\end{table}

First, in terms of the number of solved tasks, enabling or disabling intent guidance does not change the set of solvable tasks: the success rates of the two configurations are identical. This suggests that the fact-intent DAG and the task scheduling mechanism provide the foundation that enables \tool to complete long-tail tasks, and that the main contribution of the intent retrieval and prediction module is not to extend the coverage of solvable tasks but to improve the efficiency of the solving process. This result is consistent with the design intent of the paper: the fact-intent DAG is responsible for state consolidation and causal constraints and determines whether the system can maintain stable reasoning in long-horizon tasks, while the intent retrieval and prediction module provides tactical priors based on this foundation and optimizes the selection of exploration paths.

In terms of successful-task rounds, intent guidance brings a substantial efficiency gain. On medium tasks, \tool requires 7.2 rounds on average, whereas \tool--noIntent requires 10.8 rounds, a reduction of about 33\%. On hard tasks, \tool requires 9.4 rounds and \tool--noIntent 18.3 rounds, a reduction of about 48\%. Overall, the average number of rounds per successful task drops from 11 to 7. This indicates that the tactical priors provided by the intent graph through subgraph-isomorphism verification and fuzzy matching effectively guide the agent to advance preferentially along high-probability paths and avoid repeated probing of low-probability branches. The reduction is larger on hard tasks than on medium tasks, which suggests that the benefit of intent guidance becomes more pronounced as tasks become more complex and expose more alternative branches.

\begin{figure*}[!ht]
    \centering
    \includegraphics[width=0.65\linewidth]{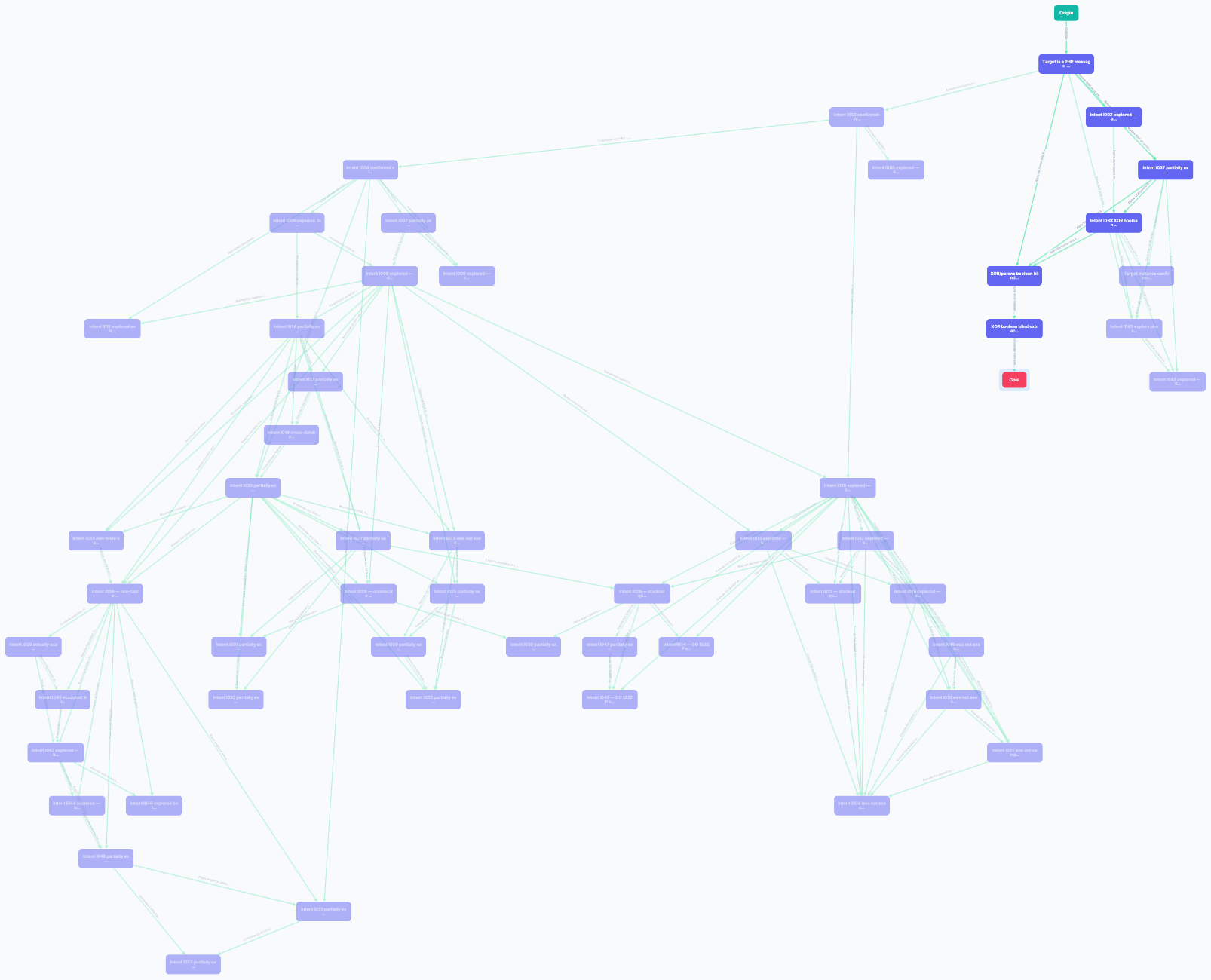}%
    \includegraphics[width=0.315\linewidth]{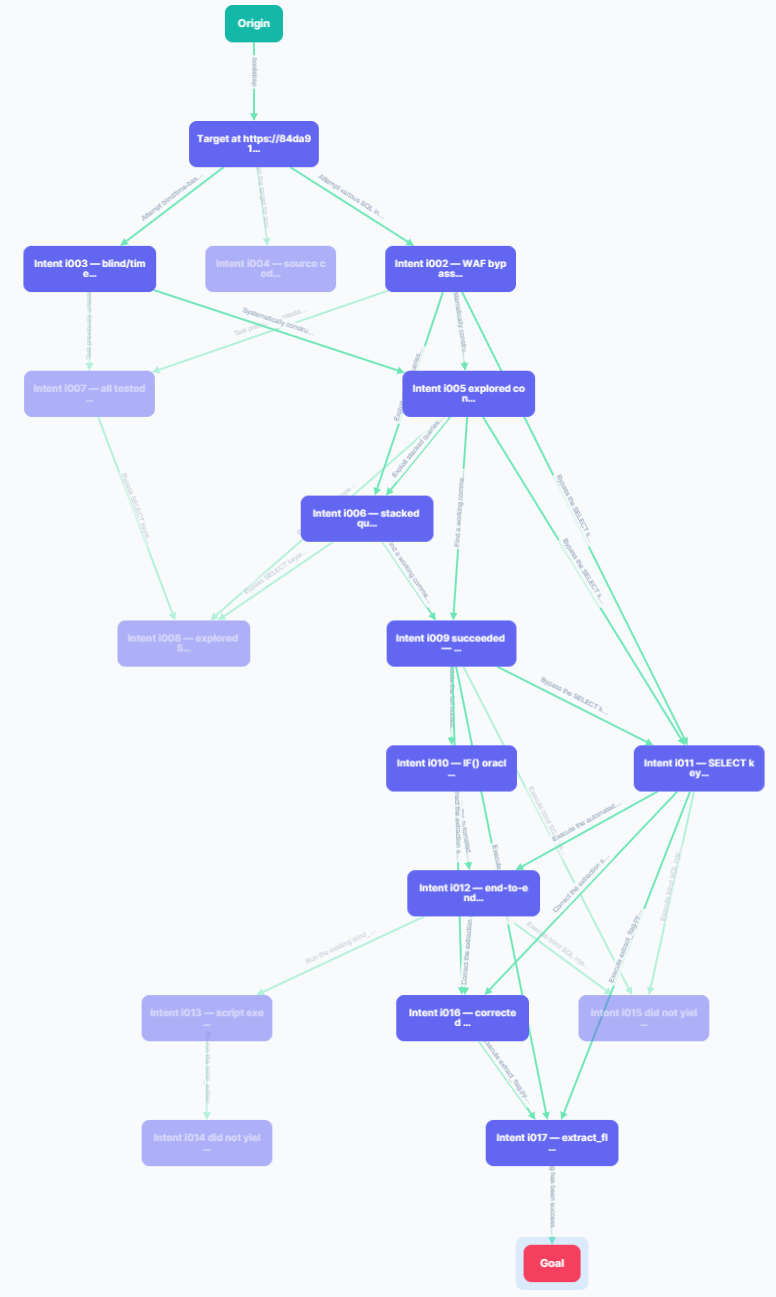}
    \caption{Exploration paths on the medium-difficulty SQL injection task: \tool--noIntent (left) vs.\ \tool (right).}
    \label{fig:medium-sql}
\end{figure*}

In terms of failed-task rounds, \tool and \tool--noIntent consume similar averages (40 vs.\ 37 rounds on medium tasks and 31.7 vs.\ 29.3 rounds on hard tasks). Intent guidance brings only a slight increase. This indicates that, while shortening successful paths, intent guidance does not markedly increase resource consumption on unsolvable tasks. Combined with the substantial reduction in successful-task rounds, the intent retrieval and prediction module reduces the overall computational cost of the system. The slight increase in failed-task rounds arises because, with intent guidance, the agent performs more targeted verification before confirming that a path is infeasible, instead of quickly falling into a loop and being reclaimed early by the scheduler when it lacks direction. This extra overhead is within an acceptable range.

In summary, the ablation results show that the fact-intent DAG provides the fundamental state consolidation and causal constraints that determine whether the system can complete long-tail tasks, while the intent retrieval and prediction module further optimizes exploration efficiency based on this foundation. Together, the two components constitute the core of the \tool design. Combined with the VulnBot comparison in RQ1, where the absence of graph-structured state coincides with a substantially lower solvable-task rate (44.1\% vs.\ 88.2\% overall), these results jointly indicate that graph state is the decisive factor for solvability within our evaluation setting, while the intent module mainly determines exploration efficiency. This evidence is examined further in Sect.~\ref{sec:discussion}.

\rqa{RQ3}{The intent retrieval and prediction module does not change the set of solvable tasks (the success rates of the two configurations are identical) but substantially improves solving efficiency: the average number of rounds per successful task drops from 11 to 7, corresponding to reductions of about 33\% on medium tasks (7.2 vs.\ 10.8) and about 48\% on hard tasks (9.4 vs.\ 18.3).}

\subsection{Case Studies}
\label{sec:case}

To illustrate the impact of intent guidance on the exploration behavior of the agent, we select two representative SQL injection cases of medium and hard difficulty and compare the exploration paths of \tool and \tool--noIntent. The quantitative ablation study above has shown that intent guidance reduces the number of rounds on successful tasks without changing the set of solvable tasks. This subsection further analyzes, from a qualitative perspective, the source of these round savings, i.e., how intent guidance helps the agent avoid low-probability branches and identify effective tactics earlier. In the path graphs of the two cases, the left side shows the exploration process of \tool--noIntent and the right side that of \tool.
Green nodes mark the origin (Origin) and red nodes the goal (Goal). The highlighted path between them is the actual attack path that reaches the goal, and the unhighlighted branches are the ineffective attempts during exploration.

\begin{figure*}[!ht]
    \centering
    \includegraphics[width=0.64\linewidth]{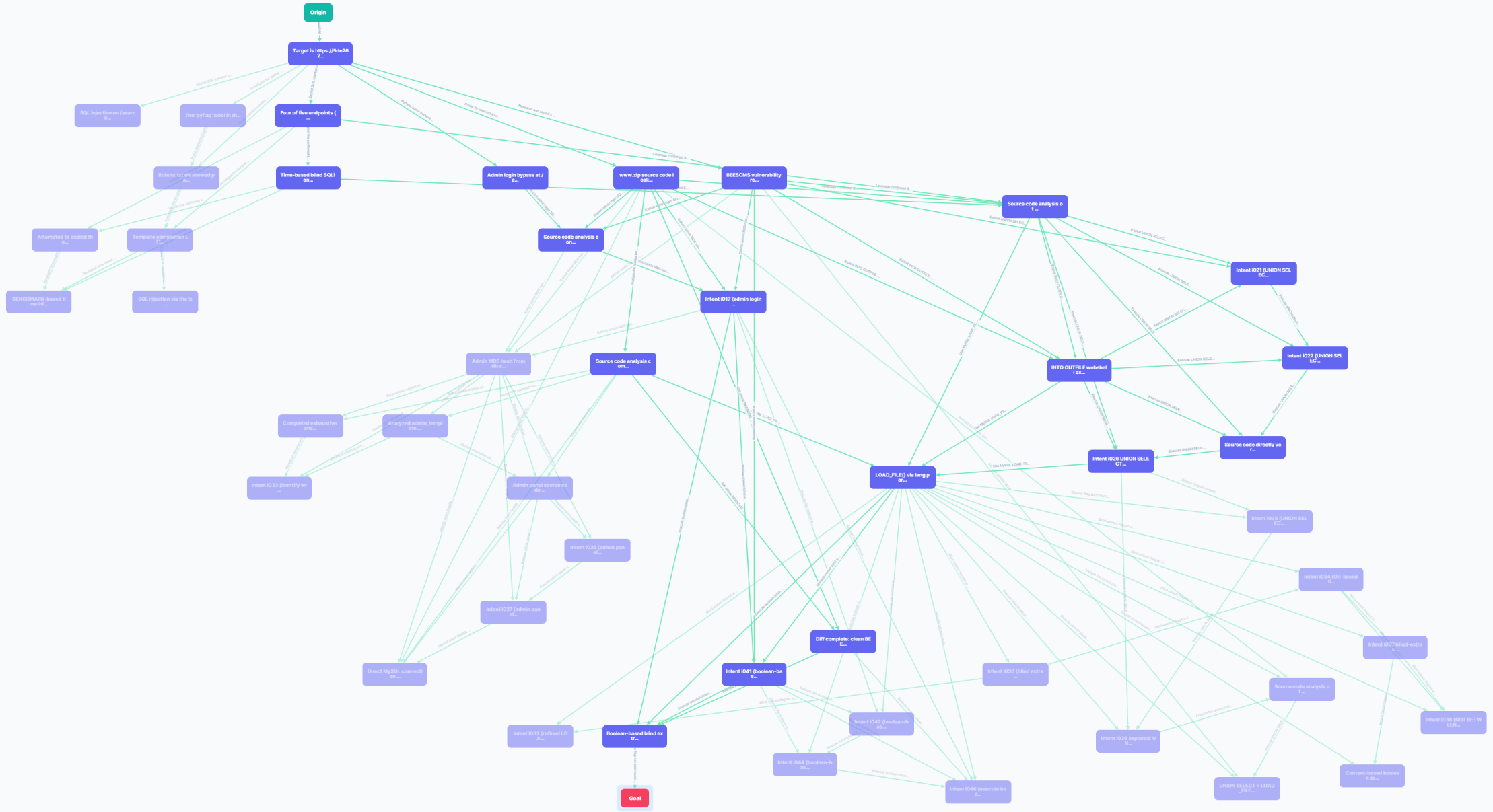}%
    \includegraphics[width=0.35\linewidth]{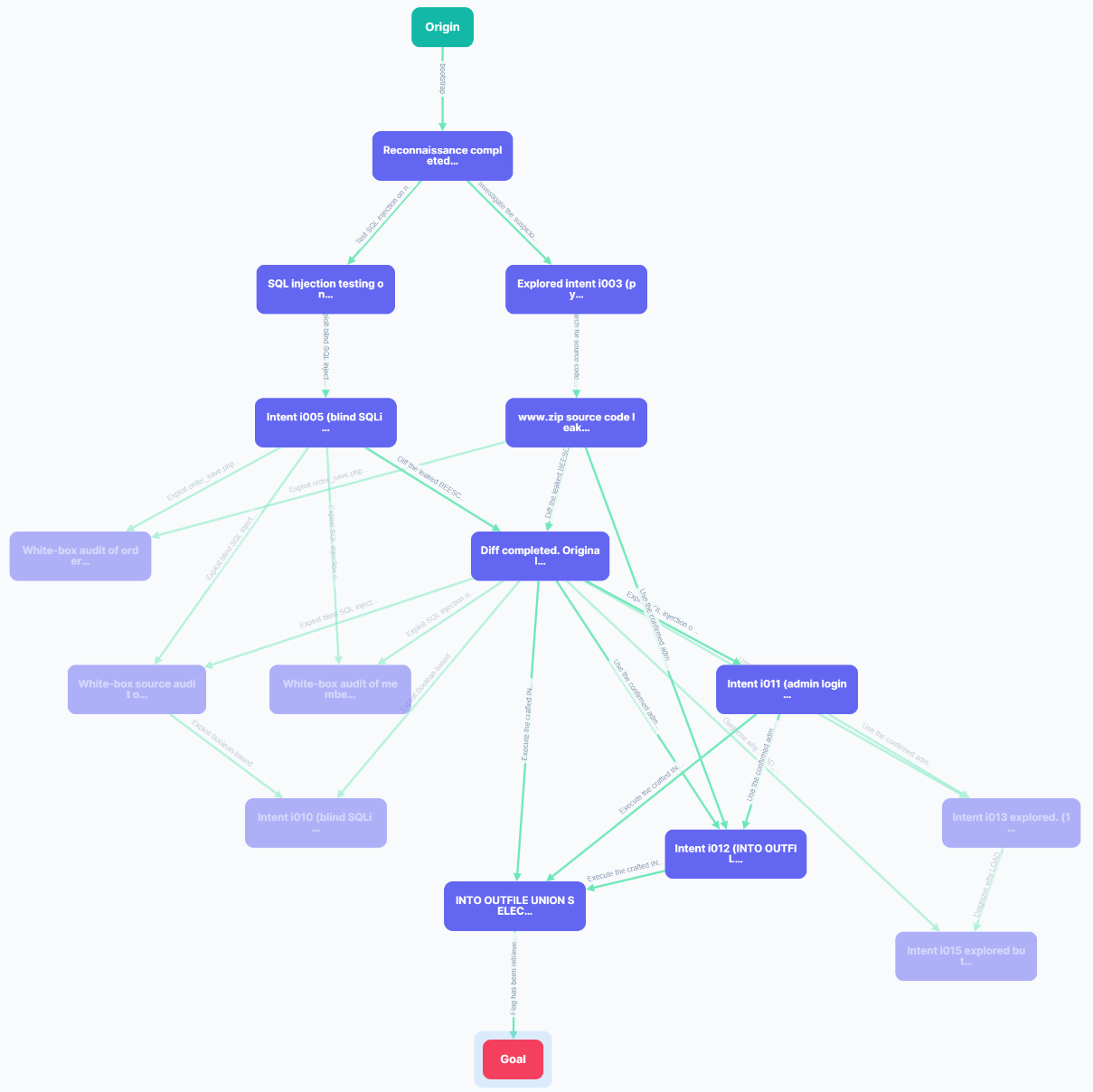}
    \caption{Exploration paths on the hard-difficulty SQL injection task: \tool--noIntent (left) vs.\ \tool (right).}
    \label{fig:high-sql}
\end{figure*}

\subsubsection{Medium-difficulty SQL injection case}

As shown in Fig.~\ref{fig:medium-sql}, this is a medium-difficulty SQL injection challenge in which the target system is protected by a web application firewall (WAF) and the agent must read a flag from the database while bypassing the WAF. In the exploration path of \tool--noIntent, without tactical priors, the agent devotes a large number of exploration rounds to the time-based injection branch. Although time-based injection yields some intermediate results in this environment, the WAF reliably detects time-delay characteristics, so the agent repeatedly encounters interception and failure on this route and must constantly adjust payloads to cope with filtering rules, consuming a large number of rounds on this low-probability branch. After many ineffective attempts, the agent finally abandons time-based injection and achieves the goal by combining XOR with Boolean-based blind injection, but the overall exploration path is markedly lengthened and contains a large number of redundant tentative operations.

With intent guidance, the exploration path of \tool converges markedly. Although the agent also tries other injection techniques such as error-based injection in the early stage, with the tactical priors provided by the intent graph through subgraph-isomorphism verification and fuzzy matching, it quickly identifies the bypass path with the highest success probability in the current environment, switches its exploration focus to that path promptly, rapidly locates the filtering defect of the WAF, bypasses it, and finally obtains the flag with considerably fewer rounds. This case shows that, when multiple feasible tactical branches exist with substantially different success probabilities, intent guidance can help the agent avoid getting stuck in local exploration on low-probability branches, thereby narrowing the search space of effective paths.

\subsubsection{Hard-difficulty SQL injection case}

As shown in Fig.~\ref{fig:high-sql}, this is a hard-difficulty SQL injection challenge built on a real web application. The target system is affected by source-code leakage, allowing its full source code to be compared with the corresponding open-source version to locate the key differentiating functions.
In the exploration path of \tool--noIntent, the agent fails to associate the early discovery of the source-code leakage with subsequent exploitation techniques. Instead, it tentatively scans a large number of pages without injection vulnerabilities, diverges in exploration direction, and spends many rounds without reaching the core attack surface.

With intent guidance, \tool quickly notices that the target system is an open-source project, associates the source-code leakage fact discovered during the probing phase with the open-source version through a diff comparison, and rapidly locates the differentiating code. It then carefully analyzes the SQL injection filtering rules in the differentiating code, tests the pages at risk in a targeted manner, successfully bypasses the SQL injection filter, writes a web shell into the web directory through INTO OUTFILE, and finally reads the flag. This case illustrates the value of intent guidance in cross-fact association: the tactical significance of source-code leakage, an early subtle fact, can be exploited only when combined with the judgment that the target is an open-source project. Without structured state association, the agent tends to treat such early facts as isolated information and ignore them. The intent graph, by connecting early facts with subsequent intents through causal links, enables the agent to trace back and exploit these facts promptly. This is precisely the role of the proposed mechanism in addressing the long-tail fact-forgetting problem.

In summary, the two cases show that intent guidance affects exploration behavior in two main ways. First, in tactical branch selection, it guides the agent to avoid low-probability branches and commit to high-probability paths earlier, reducing ineffective probing. Second, in cross-step fact association, it enables subtle facts discovered early to be traced back and integrated into subsequent reasoning promptly, avoiding fact forgetting and intent drift in long-horizon tasks. These two aspects corroborate the quantitative results of the ablation study (a substantial reduction in successful-task rounds with roughly unchanged failed-task rounds) and further explain, from a behavioral perspective, the working mechanism of the intent retrieval and prediction module.

\section{Discussion}
\label{sec:discussion}

\subsection{Failure Analysis}
\label{sec:failure}

To understand the nature of the performance gap between the baselines and \tool, we analyzed the failure logs of VulnBot, the baseline with the highest overall success rate, and identified four recurring failure patterns.
\begin{itemize}
\item \textbf{Premature success declaration (F1).} In several cases, the LLM judged a stage successful after only deriving an exploitation approach, without actually retrieving the flag. In other cases, the agent had already obtained a shell but misjudged the exploitation as failed because of erroneous LLM output and hallucinations, and took a detour that roughly doubled the number of rounds consumed. Overall, the failed tasks of VulnBot consumed about 32.8 rounds on average, below the 40-round budget, which indicates that most of its failures terminated through the agent's own judgment rather than budget exhaustion. This contrasts with Pentest-R1, whose failed-task averages of 40 indicate budget exhaustion, that is, a difference between reasoning defects and resource depletion.
\item \textbf{Long-horizon role and state confusion (F2).} On long-horizon tasks, the agent confused the attacking machine (the Kali host accessed over SSH) with the target machine and searched for the flag on the wrong host, showing a loss of the ``current host'' state over time.
\item \textbf{Missed signals in long outputs (F3).} When the flag value appeared in the middle of a long output (e.g., among many Kubernetes environment variables), the LLM omitted it when summarizing the results and reported only that ``environment-variable information was collected'' instead of recognizing that the flag had been found, which is a direct manifestation of attention dilution \cite{liu2023lost}.
\item \textbf{Rigid playbook and redundant re-execution (F4).} Regardless of the actual target, the agent first ran the preset scanners, attempted SQL injection whenever parameters were observed without judging whether injection was possible, invoked sqlmap for every SQL-injection scenario, and did not actively bypass the WAF. In addition, stages already completed were re-executed in later planning, and as the task grew longer, the agent questioned flags it had already obtained, re-verified them, and produced hallucinated false alarms.
\end{itemize}
Taken together, these patterns point to a common cause: fixed procedural patterns that adapt poorly to the specific scenario, plus unreliable state memory that leads to repeated verification and self-doubt, rather than budget exhaustion. The four patterns correspond to the three problems analyzed in the motivation section: F1, F2, and F3 to the first problem, context forgetting and intent drift in long-horizon tasks, and F4 to the third problem, the limitations of retrieval mechanisms and static fine-tuning, under which rigid procedural patterns substitute for adaptive tactical priors. The second problem, runtime reasoning and operation-boundary control risks, concerns mechanism-level defenses that do not surface in the task-level failure logs of the baselines. However, F2 still reflects this risk to some extent, as the agent attempts to execute potentially risky commands on the wrong host.

The few tasks where \tool failed exhibit a different failure profile. The hard tasks on which \tool failed are expert-level challenges that combine multiple vulnerabilities into a single exploitation chain and require framework-internal knowledge to solve. A representative case requires bypassing a path-parsing authentication defect of a web framework in combination with a filter-bypass technique for a database driver's initialization mechanism, and even the official solution demands local debugging and source-level inspection of the involved frameworks, so that human experts also consider such tasks highly complex. Therefore, the failure reflects the difficulty of the task rather than a defect in reasoning or state management. In the case of the medium task where \tool failed, the agent had already located the file containing the flag, but the final step required privilege escalation on the target server, which failed, so the flag file could not be read. The failure occurred at the last privilege-escalation step rather than in discovery or reasoning.

Taken together, these observations reveal two distinct failure profiles. The failures of VulnBot are reasoning and state defects: premature success declaration (F1), role and state confusion (F2), missed signals in long outputs (F3), and rigid playbooks with redundant re-execution (F4), and none of them result from budget exhaustion, as reflected by its failed-task average of about 32.8 rounds against a 40-round budget. The failures of \tool are of a different nature: they concentrate on expert-level tasks that combine multiple vulnerabilities into one exploitation chain and require framework-internal knowledge, where the failure occurs at the difficulty ceiling of the task rather than at the reasoning or state layer. This contrast indicates that the state and reasoning mechanisms of \tool substantially reduce the failure modes that dominate the baselines.

\subsection{Evidence for the Necessity of Graph State}

Because the DAG is the architectural substrate of \tool, a controlled ablation that removes the graph while keeping the rest of the system intact is not feasible. We therefore assemble three lines of evidence. First, the \tool--noIntent ablation retains the DAG and the scheduler and shows that the solvable-task set is preserved, which attributes solvability to the DAG and the scheduling mechanism and attributes efficiency to the intent module. Second, VulnBot provides a cross-system reference for the absence of graph state: it is a multi-agent framework that shares plain-text test memory without graph-structured state, and its overall success rate of 44.1\% (vs.\ 88.2\% for \tool) and its failure patterns F1--F4 (Sect.~\ref{sec:failure}), which are all symptoms of unreliable state memory, approximate how a system without graph state behaves on the same benchmark. Third, the failure analysis shows that the failures of the baselines concentrate on state confusion and repeated verification, which are precisely the defects the DAG is designed to prevent.

These three lines of evidence jointly indicate that graph-based state is the decisive factor for solvability and that plain-text memory limits VulnBot, the baseline with the highest overall success rate. This triangulation is corroborating rather than conclusive, since VulnBot differs from \tool in orchestration and tooling as well. The residual attribution uncertainty is stated in Sect.~\ref{sec:threats}.

\subsection{Role and Applicability Boundary of the Intent Graph}
RQ3 shows that disabling intent retrieval does not change the set of solvable tasks, which indicates that the intent graph acts mainly as a structured tactical prior rather than as a source of new attack knowledge. This observation defines the applicability boundary of the intent graph. On attack surfaces with known or similar historical patterns, the graph prunes low-probability branches and accelerates the agent. On attack surfaces without any historical features, the system gains no exploitable prior knowledge from the graph. Its benefit there comes from the constraint and scheduling guarantees of the DAG, while the LLM's heuristic reasoning (Stage 5 of the retrieval pipeline) preserves the ability to explore unknown environments under the precondition validation described in Sect.~\ref{sec:retrieval}. The quantitative manifestation of this boundary is the RQ3 result: the solvable-task set is unchanged, while the efficiency gains are substantial. We interpret this as a division of labor rather than as a limitation: the DAG determines whether the agent can sustain long-horizon reasoning, and the intent graph determines how efficiently it explores.

The offline intent graph has inherent timeliness and coverage gaps. It is built from historical reports and exercise documents, so new vulnerability classes, 0-day variants, and attack patterns that have not appeared in the corpus are absent from the graph. For such attack surfaces, the first four retrieval stages cannot contribute priors, and the system operates through Stage 5 under DAG constraints. This is not a behavioral degradation of the system, but it does mean that the quality of the tactical priors depends on the freshness of the graph. As a direction for future work, the graph can be updated dynamically: once new attack patterns are verified at runtime (that is, an intent chain that succeeds in the live environment), they can be incorporated into the graph as new intent edges, so that the prior knowledge grows with the deployment history of the system.

\section{Threats to Validity}
\label{sec:threats}

\paragraph{Internal validity}
All agents in our experiments invoke the same LLM, DeepSeek V4 Pro, through its API, which controls for model-level confounds across methods. However, the configuration of the baselines may introduce variance: Claude Code, a general-purpose agent, requires a task-specific prompt and the ECC Skills plugin, and the exact prompt design may influence its performance. PentestGPT depends on a human expert to execute its suggested operations, and the expert's operational style may vary. In addition, the stochastic nature of LLM inference may add noise to the round counts reported in this study.

\paragraph{External validity}
The benchmark is built from records of real CTF competitions, and the challenges involve vulnerabilities of real web application systems. Nevertheless, several gaps separate this setting from production environments. First, the benchmark consists of a limited set of CTF challenges, far smaller than the asset scale of production networks, and the coarse three-level difficulty binning may not capture the full spectrum of task complexity. Second, CTF challenges typically involve single applications or small network segments, whereas production enterprise intranets contain large-scale topologies and heterogeneous services, where intranet traversal and cross-segment lateral movement play a central role. Third, the defense mechanisms embedded in CTF challenges, such as WAFs and honeypots, differ from the dynamic defenses of production environments, such as EDR, IDS, micro-segmentation, and dynamically changing network policies, which may alter the cost and success probability of specific attack steps. Fourth, production testing is further constrained by business continuity and compliance requirements: aggressive operations that are acceptable in a CTF environment, such as repeated brute-forcing or privilege escalation attempts, are restricted in production engagements, so the operational envelope of the agent in practice differs from the benchmark. Finally, the size of the benchmark is constrained by the data and computational resources available for this study, and expanding it to a wider set of scenarios remains future work.
In summary, the reported results support the effectiveness of \tool within web-based CTF environments. The generalization of \tool to production enterprise intranet environments and to vulnerability types and protocols not covered by the benchmark remains to be validated in future work.

\paragraph{Construct validity}
Task success rate and the average number of rounds per task are standard and objective metrics, but they are coarse proxies for the effectiveness and efficiency of penetration testing agents. Other dimensions such as the severity of the exploited vulnerabilities, the practicality of the obtained access, or the time and API cost of the agents are not directly measured. The round counts of failed tasks reflect the exploration performed until termination, which is affected by the termination policy of the scheduler and the task budget.

\paragraph{Ablation design} A fully controlled ablation of the fact-intent DAG itself is not feasible: the DAG is the architectural substrate of \tool, and removing it would amount to rebuilding the system rather than disabling a module. We therefore provide evidence for the necessity of graph state through three complementary routes: (1) the \tool--noIntent ablation, which isolates the contribution of the intent retrieval module while retaining the DAG. (2) The comparison with VulnBot, a multi-agent system with plain-text shared memory but no graph state, whose markedly lower success rate (44.1\% vs.\ 88.2\% overall, 25.0\% vs.\ 75.0\% on hard tasks) and failure patterns (Sect.~\ref{sec:failure}) approximate the behavior of a system without graph state. (3) The failure-pattern analysis, which traces the failures of the baselines to state and reasoning defects that the DAG is designed to prevent. We acknowledge that the VulnBot comparison is a cross-system contrast rather than a controlled ablation: the two systems also differ in orchestration structure and tooling, so this evidence is corroborating rather than conclusive. The corresponding discussion is presented in Sect.~\ref{sec:discussion}.

\paragraph{Data contamination and leakage} The offline intent graph is built by an LLM-driven extractor from public penetration testing reports, threat intelligence, and exercise documents, which raises the risk of evaluation leakage if material related to the test challenges enters the graph.
To control this risk, the challenges used in our experiments and their publicly available writeups were excluded from the extraction corpus through a blacklist covering challenge names, originating competitions, and writeup repositories, and the remaining corpus was verified by sampling checks (Sect.~\ref{sec:setup}). In addition, the internal validator of the extraction pipeline discards extracted graphs whose confidence scores fall below a set threshold. A residual risk remains: paraphrased or variant writeups that do not match the blacklist keywords could still introduce test-related patterns, and this risk cannot be fully eliminated for public challenge material.

\paragraph{Scalability} The graph storage, retrieval, and matching mechanisms of \tool were evaluated in CTF scenarios, where the runtime attack graph remains small. Applying graph matching to production-scale networks raises a known scalability concern. The design of \tool mitigates this concern: subgraph isomorphism is applied only to the local candidate subgraphs retained by the Top-$k$ pre-screening rather than to the whole graph, and the fuzzy Jaccard fallback avoids hard isomorphism failures, but the end-to-end scalability of graph storage, retrieval, and matching on very large networks has not been measured in this study and remains to be validated. All agents invoked DeepSeek V4 Pro through its API, and the systematic cost analysis at scale is left to future work.

\paragraph{Adversarial contamination and security boundary} The security boundary of \tool rests on the structural precondition validation of the DAG and container-level isolation (Sect.~\ref{sec:sched}), but the residual risks deserve a threat-modeling discussion. (a) Offline intent-graph poisoning: malicious threat-intelligence reports or exercise documents in the extraction corpus could introduce incorrect attack patterns into the offline graph. The confidence threshold and validation of the extractor mitigate but do not eliminate this risk. (b) Retrieval-corpus pollution: adversarial material that resembles legitimate reports could degrade the quality of the tactical priors retrieved at runtime. (c) Runtime injection of malicious intent edges: an agent manipulated by indirect prompt injection could attempt to write intents that are not grounded in verified facts. Such intents are rejected by precondition validation and cannot obtain execution authorization, with container isolation as a further containment layer. Under these mitigations, the remaining risks are confined to degraded decision quality rather than unauthorized execution. Robustness evaluations under adversarial graph manipulation are left to future work.

\section{Conclusion and Future Work}
\label{sec:conclusion}

This paper proposes \tool, an intent-graph-guided automated penetration testing agent, and effectively mitigates the context forgetting and intent drift of LLMs in long-tail penetration testing tasks. The method employs a fact-intent DAG as the persistent state source: verified network states are consolidated as immutable fact nodes, and exploration directions are constrained as intent edges governed by predecessor facts, mitigating aimless trial-and-error and context collapse.
A task scheduling mechanism with two-phase degradation recovery and multi-dimensional adaptive load balancing supports execution stability, and a top-down five-stage intent retrieval and prediction algorithm provides tactical priors. On a real CTF test set covering multiple mainstream web vulnerabilities with a difficulty gradient, \tool achieves an overall task success rate of 88.2\% and a success rate of 75.0\% on hard tasks, improving by approximately 44 and 50 percentage points over VulnBot, the baseline with the highest overall success rate. The ablation study shows that the intent retrieval and prediction module reduces the average number of rounds on successful medium and hard tasks by about 33\% and 48\%, respectively, without changing the set of solvable tasks.
The experimental results corroborate the effectiveness of the fact-intent DAG in suppressing context forgetting and intent drift, as well as the role of intent guidance in shortening exploration paths. Future work includes extending the intent graph to more vulnerability types with a dynamic online update mechanism, evaluating the system in larger-scale network environments beyond CTF challenges, and investigating the transferability of the fact-intent DAG to other long-horizon cybersecurity agent tasks.

\section*{Acknowledgment}
This work is supported in part by Hainan Province Science and Technology Special Fund (Grant No. ZDYF2024GXJS008), in part by the National Science Foundation of China under Grants U22B2027, U2468204, and U25A20424, the Joint Special Project of Beijing-Tianjin-Hebei Natural Science Foundation under Grant 25JJJJC0033, the Tianjin Special Fund Project for High-quality Development of Manufacturing Industry under Grant 20251148, and Tianjin Science and Technology Plan Project under Grant 23YDPYGX00140.
We also thank the authors and contributors of Cairn for their open-source project, which provided valuable support for this work.

\bibliographystyle{elsarticle-num}
\bibliography{reference}   

\end{document}